\documentclass[jcp,twocolumn,showpacs,superscriptaddress,10pt]{revtex4}
\usepackage{graphicx}
\usepackage{subfigure}
\usepackage{amsmath}
\usepackage{amsfonts}
\usepackage{amssymb}
\usepackage{amsthm}
\usepackage{times}
\usepackage[colorlinks,citecolor=blue,linkcolor=blue]{hyperref}
\usepackage{color}

\begin{document}
\title{Nonequilibrium spin-boson model: from weak to strong coupling}

\newcommand{\Fudan}{\affiliation{State Key Laboratory of Surface
Physics and Department of Physics, Fudan University, Shanghai 200433,
China}}
\newcommand{\Boulder}{\affiliation{Department of Mechanical Engineering, University of Colorado, Boulder, CO 80309, USA}}
\newcommand{\CI}{\affiliation{Collaborative Innovation Center of Advanced Microstructures, Fudan University, Shanghai 200433, China}}

\author{Junjie Liu}
\Fudan
\author{Hui Xu}
\Fudan
\author{Baowen Li}
\Boulder
\author{Changqin Wu}
\email{cqw@fudan.edu.cn}
\Fudan
\CI

\begin{abstract}
We present a general theory to explore energy transfer in nonequilibrium spin-boson models within the framework of nonequilibrium Green's function (NEGF). In contrast to conventionally used NEGF methods based on a perturbation expansion in the system-bath coupling, we adopt the polaron transformation to the Hamiltonian and identify the tunneling term as a perturbation with the system-bath coupling being treated nonperturbatively, herein termed the polaron-transformed NEGF method. To evaluate terms in the Dyson series, we further utilize the Majorana-fermion representation. The proposed method not only allows us to deal with weak as well as strong coupling regime, but also enables an investigation on the role of bias. As an application of the method, we study the energy transfer between two Ohmic bosonic baths mediated by a spin. For a unbiased spin system, our energy current result smoothly bridges predictions of two benchmarks, namely, the quantum master equation and the nonequilibrium non-interacting blip approximation, thus our method is beyond existing theories. In case of a biased spin system, we reveal a bias-induced nonmomotonic behavior of the energy conductance in the intermediate coupling regime, due to the resonant character of the energy transfer. This finding may offer a nontrivial quantum control knob over energy transfer at the nanoscale.
\end{abstract}

%05.60.Gg:quantum transport
%05.30.-d:quantum statistical mechanics
%44.10.+i:heat conduction

\maketitle

\section{Introduction}
Recently, energy transfer at the nanoscale has received
significant attention and has grown considerably in importance. From an experimental perspective, several procedures have been developed to measure energy transfer at a microscopic level, such as the time-domain thermoreflectance techniques \cite{Koh.07.PRB},
the scanning thermal microscopy \cite{Lefevre.05.PSI} and the laser Raman
scattering thermometry \cite{Chavez.14.APLM,Reparaz.14.PSI}. Such experimental advances stimulated a surge of theoretical activity in understanding and controlling energy transfer in nanoscale conductors \cite{Wang.08.EPJB,Dubi.11.RMP}. In particular, the energy transfer in the nonequilibrium spin-boson model (NESB) is now an established area of theoretical research \cite{Segal.05.PRL,Ren.10.PRL,Boudjada.14.JPCA}.

To date, various approaches have been carried out to deal with energy transfer in the NESB. On the theoretical side, existing methods are developed along two main lines, depending on which term in the Hamiltonian is identified for the perturbation expansion. One line, treating the system-bath coupling in a perturbative manner, includes two eminent methods: the quantum master equation (QME)
analysis \cite{Segal.05.PRL,Segal.06.PRB,Ruokola.11.PRB,Thingna.14.JCP} and the nonequilibrium Green's function (NEGF) technique \cite{Velizhanin.10.JCP,Wang.13.FP,Yang.14.EL}. Within the Redfield approximation the QME can describe incoherent sequential transfer process \cite{Segal.05.PRL,Segal.06.PRB} and even possible to include cotunneling process if the generalized Fermi golden rule is utilized \cite{Ruokola.11.PRB}. Although simple and physically transparent analytical results are obtained, the QME is restricted to weak coupling regime. The NEGF provides an alternative, it hands over a formally exact expression for energy flux written in terms of the spin correlation functions (GFs). A perturbation expansion in the system-bath coupling is adopted only in obtaining those GFs. Velizhanin and coworkers \cite{Velizhanin.10.JCP} work out expressions for GFs using the Redfield approximation, however their results violate the energy conservation. In order to preserve conservation laws, a Majorana-Fermion diagrammatic method (MFDM) \cite{Mao.03.PRL,Shnirman.03.PRL,Florens.11.PRB,Schad.15.AP} is further used in calculating GFs \cite{Yang.14.EL}. Compared with the QME, the NEGF can be applied to the intermediate coupling regime, but the strong coupling regime is still beyond its scope.

The other line chooses the tunneling term as a perturbation, being followed by the nonequilibrium version of the non-interacting blip approximation (NE-NIBA) \cite{Nicolin.11.JCP,Nicolin.11.PRB} and a nonequilibrium polaron-transformed
Redfield equation (NE-PTRE) method \cite{Wang.15.SR}. However, the NE-NIBA is valid only in the strong coupling regime \cite{Chen.13.PRB}. The NE-PTRE, although provides a correct picture for the NESB with super-Ohmic environments, reduces to the NE-NIBA framework for the NESB with Ohmic or sub-Ohmic ones \cite{Wang.15.SR} and thus face the same pathology.

So far, a systematic investigation on the energy transfer process from weak to strong system-bath couplings for the NESB embedded in Ohmic bosonic baths is still absent \cite{Boudjada.14.JPCA}.
To go beyond perturbation theories, exact numerical techniques have been proposed, like multilayer
multiconfiguration time-dependent Hartree theory \cite{Velizhanin.08.CPL}, influence functional path integral techniques \cite{Segal.13.PRB} and classical Monte Carlo simulations \cite{Saito.13.PRL}. However, numerical simulations can become time-consuming in some parameter regimes and sometimes rely on exact mappings between different systems, which limits their applicability.

Furthermore, from both a theoretical and numerical point of view, little is known about the role of a finite bias plays in the energy transfer of NESB with Ohmic dissipations, only very recently, the time-dependent energy transfer under the condition of a time-dependent bias is studied within an influence functional approach \cite{Carrega.16.PRL}. However, the steady state properties still remain unknown, especially at the intermediate coupling regime \cite{Boudjada.14.JPCA}. Recalled that the Ohmic bath serve as a paradigm for the simulation of environments with an abundance of low frequency modes, such as liquids, proteins,
and polymers \cite{Makarov.94.CPL}, it is therefore of a great importance to introduce a systematic approach that allows for an arbitrary system-bath coupling, handles finite bias, to study energy transfer in such systems.

In this work, for the first time, we fill the gap and propose a unified theoretical scheme which is capable of bridging results of two benchmarks, namely, the QME and the NE-NIBA in the unbiased system and meanwhile, including effects of a finite bias. We note that a combination of the MFDM and the NEGF leads to conservation laws and can also be applied to a wide range of temperature in the weak coupling regime \cite{Yang.14.EL,Bijay.16.NULL}. Thus we follow this framework and absorb these advantages into our theory. However, in contrast to conventional NEGF methods treating the system-bath interaction in a perturbative manner, we make the polaron transformation (PT) \cite{Weiss.12.NULL} on the Hamiltonian and perform a perturbation expansion in the tunneling term. Since our perturbation expansion is nonperturbative in the system-bath coupling, we can explore the strong system-bath coupling regime as we have demonstrated in equilibrium systems \cite{Liu.16.CP}. Our theory can be regarded as an extension to the previous NEGF methods, thus we refer the proposed scheme to "PT-NEGF" method.

With the PT-NEGF, we investigate the energy transfer of the NESB embedded in the Ohmic bosonic baths in details. For unbiased spin systems, our energy current result smoothly bridges results of the QME and the NE-NIBA. In case of finite bias, we reveal a nonmomotonic bias dependence of the conductance in an intermediate coupling regime with moderate temperatures. We attribute this dependence to the resonant character of the heat transfer. These features thus make our theory stand out from previous considerations.

The paper is organized as follows. We first introduce the NESB model and its non-equilibrium environment in section \ref{sec:2}. In section \ref{sec:3}, we present methodologies of the PT-NEGF. In section \ref{sec:4}, we study the energy transfer of the NESB using the PT-NEGF in detail. In section \ref{sec:5}, we summarize our findings and make some final remarks.

\section{Backgrounds}\label{sec:2}
\subsection{Model system}
The nonequilibrium spin boson (NESB) model, consisting of a two level system in contact with two bosonic reservoirs, is described by the general Hamiltonian \cite{Leggett.87.RMP}
\begin{equation}\label{eq:h}
H~=~H_s+H_I+H_B,
\end{equation}
where the system Hamiltonian $H_s=\frac{\varepsilon}{2}\sigma_z+\frac{\Delta}{2}\sigma_x$ with $\varepsilon$ the bias, $\Delta$ the tunneling between two levels and $\sigma_{x,z}$ the Pauli matrices. Since the spectrum of the system Hamiltonian is symmetric for positive and negative bias, here we only consider positive bias. The bath part $H_B=\sum_{v=L,R}H_B^v=\sum_{j,v=L,R}\omega_{j,v}b^{\dagger}_{j,v}b_{j,v}$ and the bilinear interaction term $H_I=\sigma_z\sum_{j,v=L,R}g_{j,v}(b^{\dagger}_{j,v}+b_{j,v})$ with
$b^{\dagger}_{j,v}(b_{j,v})$ the creation (annihilation) operator of the $j$th harmonic mode in the $v$ bosonic bath and $g_{j,v}$ the system-bath coupling strength. Throughout the paper, we set $\hbar=1$ and $k_B=1$. The influence of the bath
is contained in the spectral density function
\begin{equation}
I_v(\omega)=2\pi\sum_{j\in v}g_{j,v}^2\delta(\omega-\omega_{j,v}).
\end{equation}
For convenience, we make the specific choice \cite{Weiss.12.NULL}
\begin{equation}\label{eq:bath_spec}
I_v(\omega)~=~\pi\alpha_v\omega^s\omega_{c}^{1-s} e^{-\omega/\omega_{c}}.
\end{equation}
where $\alpha_v$ is the dimensionless system-bath coupling strength between the $v$ bath and the spin system, $\omega_c$ is the cutoff frequency of the bath (We choose the same cutoff frequency for two baths). The case $s>1(s<1)$ corresponds to super-Ohmic (sub-Ohmic) dissipation, and $s=1$ represents the important case of frequency-independent (Ohmic) dissipation. In this study, we limit ourselves to Ohmic dissipations.

\subsection{Definition of energy current and energy conductance}
We utilize the following definition for the energy current from $v$th bosonic bath to the system
\begin{equation}
J_{v}~=~-\frac{\partial}{\partial t}\langle H_B^{v}\rangle
\end{equation}
with $H_B^v=\sum_{j,v}\omega_{j,v}b^{\dagger}_{j,v}b_{j,v}$ the Hamiltonian of $v$th bosonic bath. It is worthwhile to mention that the above definition is consistent with the quantum thermodynamics and can be applied to the strong coupling regime \cite{Esposito.15.PRL}.

We introduce Green's functions (GFs) of Pauli matrices on the Keldysh contour \cite{Rammer.86.RMP,Haug.96.NULL}
\begin{equation}
\Pi_{\alpha\beta}(t,t^{\prime})~=~-i\left\langle T_c \sigma_{\alpha}(t)\sigma_{\beta}(t^{\prime})\right\rangle ,\qquad\alpha,\beta=x,y,z,
\end{equation}
where $T_c$ is the
contour-ordered operator responsible for the rearrangement of operators according to their contour time. The earlier (later) contour time places operators to the right (left). Its retarded, advanced, lesser and greater components are give by
$\Pi_{\alpha\beta}^r(t,t^{\prime}) = -i \Theta(t-t^{\prime})\langle [\sigma_{\alpha}(t),\sigma_{\beta}(t^{\prime})]\rangle$,
$\Pi_{\alpha\beta}^a(t,t^{\prime}) = i \Theta(t^{\prime}-t)\langle [\sigma_{\alpha}(t),\sigma_{\beta}(t^{\prime})]\rangle$,
$\Pi_{\alpha\beta}^<(t,t^{\prime}) = -i\langle\sigma_{\beta}(t^{\prime})\sigma_{\alpha}(t)\rangle$ and
$\Pi_{\alpha\beta}^>(t,t^{\prime}) = -i\langle\sigma_{\alpha}(t)\sigma_{\beta}(t^{\prime})\rangle$,
respectively, where $\Theta(\tau)$ denotes the Heaviside step function and
$\sigma_{\alpha}(\tau)\equiv e^{i H\tau}\sigma_{\alpha}e^{-i H\tau}$
denotes the operator in the Heisenberg picture.

Noting that in nonequilibrium steady states, the Keldysh GFs depends only on the difference in time, $t-t^{\prime}$. Therefore, in terms of Keldysh GFs defined above, a formal expression for the energy current from $v$th bosonic bath to the system can be expressed as \cite{Ojanen.08.PRL,Saito.13.PRL,Yang.14.EL}
\begin{equation}\label{eq:SBJ}
J_v ~=~ \frac{1}{2\pi}\int_{0}^{\infty}d\omega\omega I_v(\omega)\left[2n_v(\omega)\tilde{\chi}^{\prime\prime}_z(\omega)+\mathrm{Im}\Pi_{zz}^<(\omega)\right],
\end{equation}
where $\Pi_{zz}^<(\omega)$ is the Fourier transform of the lesser Green's function $\Pi_{zz}^<(t-t^{\prime})$, "$\mathrm{Im}$" denotes the imaginary part, and $n_v$ is the Bose-Einstein distribution of temperature $T_v$.  $\tilde{\chi}^{\prime\prime}_z(\omega)$ represents the imaginary part of the dynamical susceptibility, it is given by $-\mathrm{Im}\Pi_{zz}^r(\omega)$, or equivalently, $\left[\Pi_{zz}^<(\omega)-\Pi_{zz}^>(\omega)\right]/2i$.

By considering zero dimensionality of the system and the conservation law of current $J_L+J_R=0$, the above energy current formula [Eq. (\ref{eq:SBJ})] can be rewritten into a Landauer-type form \cite{Saito.13.PRL}
\begin{equation}\label{eq:heat_current}
J_L~=~\frac{\alpha\xi}{4\pi}\int_0^{\infty}d\omega\omega\tilde{\chi}^{\prime\prime}_z(\omega)\tilde{I}(\omega)[n_L(\omega)-n_R(\omega)],
\end{equation}
where $\alpha=\alpha_L+\alpha_R$, $\xi=4\alpha_L\alpha_R/\alpha^2$ is an asymmetry factor, for Ohmic dissipations, we have $\tilde{I}(\omega)=\pi\omega e^{-\omega/\omega_{c}}$. In the linear response regime, the energy conductance defined by $\kappa\equiv \left.d J_L/d T_L\right|_{T_L\rightarrow T_R=T}$ is given by \cite{Saito.13.PRL}
\begin{equation}\label{eq:heat_conductance}
\kappa~=~\frac{\alpha\xi}{4\pi}\int_0^{\infty}d\omega\left.\tilde{\chi}^{\prime\prime}_z(\omega)\right|_{T_L= T_R=T}\tilde{I}(\omega)\left[\frac{\omega/2T}{\sinh(\omega/2T)}\right]^2.
\end{equation}
From Eqs. (\ref{eq:heat_current}) and (\ref{eq:heat_conductance}), we know that $\tilde{\chi}^{\prime\prime}_z(\omega)$ totally determines the transport properties of the NESB. In the following section, we will develop the PT-NEGF method to obtain its explicit expression.

\section{PT-NEGF}\label{sec:3}
\subsection{Polaron transformation}
Since $\tilde{\chi}^{\prime\prime}_z(\omega)$ is of primary interest , we will mainly focus on the calculation of the Keldysh GFs of $\sigma_z$ in the following, but the methodologies discussed below carry over easily to other Keldysh GFs as well with cautions \cite{Schad.16.PRB}. We principally work here in the so-called nonadiabatic limit of $\Delta/\omega_c\ll1$. For such fast baths, it has been demonstrated that the PT is suitable for the entire range of system-bath coupling strength \cite{Lee.12.JCP}. Thus we make the PT with the unitary operator \cite{Weiss.12.NULL,Lee.12.JCP}
\begin{equation}\label{eq:U}
U=\mathrm{exp}[i\sigma_z\Omega/2],\qquad\Omega=2i\sum_{j,v}\frac{g_{j,v}}{\omega_{j,v}}(b^{\dagger}_{j,v}-b_{j,v})
\end{equation}
on the Hamiltonian Eq. (\ref{eq:h}) such that
\begin{equation}
H_T ~=~ U^{\dagger}HU= \tilde{H}_0+\tilde{H}_I,
\end{equation}
where the total free Hamiltonian is $\tilde{H}_0=\tilde{H}_s+\tilde{H}_B$ with the transformed system Hamiltonian reads
\begin{equation}
\tilde{H}_s=\frac{\varepsilon}{2}\sigma_z,
\end{equation}
and the bath Hamiltonian remains unaffected, $\tilde{H}_B=H_B$. The transformed interaction term, originating from the tunneling term in Eq. (\ref{eq:h}), takes the following form
\begin{equation}\label{eq:t_interaction}
\tilde{H}_I~=~\frac{\Delta}{2}(\sigma_x\cos\Omega+\sigma_y\sin\Omega).
\end{equation}
It's evident that $\tilde{H}_I$ contains arbitrary orders of the system-bath coupling strength by noting the form of $\Omega$ in Eq. (\ref{eq:U}).

In order that a perturbation theory can be developed for $H_T$, $\langle \tilde{H}_I\rangle_{\tilde{H}_0}=0$ should be fulfilled \cite{Lee.12.JCP}. For the bath spectral function we choose [c.f., Eq. (\ref{eq:bath_spec})], it can be easily verified that for $s\leqslant 1$ the expectation of $\tilde{H}_I$ will always approach to zero regardless of the system-bath coupling strength, while for $s>1$, the expectation is finite \cite{Wang.15.SR}. Therefore, in case of Ohmic dissipations we consider here, $\tilde{H}_I$ can be safely treated as a perturbation. The extension to the super-Ohmic dissipation is quite straightforward, we should adopt a fluctuation-decoupling scheme \cite{Lee.12.JCP,Wang.15.SR} and choose $\tilde{H}_I-\langle \tilde{H}_I\rangle_{\tilde{H}_0}$ as the transformed interaction term. By doing so, we follow a totally different perspective to develop a NEGF method, namely, the theory is perturbative in tunneling and nonperturbative in system-bath coupling.

Furthermore, we note that $[\sigma_z, U]=0$, implying expressions for the energy current as well as the heat conductance [c.f., Eqs. (\ref{eq:heat_current}) and (\ref{eq:heat_conductance})] are invariant under the PT. Thereby after the PT, we still only need to evaluate the Keldysh GFs of $\sigma_z$ with respect to the transformed Hamiltonian $H_T$.

\subsection{Majorana-fermion representation}
Note that spin operators do not satisfy the Wick theorem. In order to overcome this difficulty, the so-called Majorana-Fermion representation (MFR) \cite{Tsvelik.92.PRL} is utilized in the method such that standard Feynman diagram techniques as well as the Dyson's equation can be used. Technically, the MFR involves the introduction of a triplet of real fermions $\eta_{\alpha}$ (with $\alpha = x,y,z$) that satisfy \cite{Shnirman.03.PRL}
\begin{equation}
\eta_{\alpha}\eta_{\beta}~=~-\eta_{\beta}\eta_{\alpha}~~(\alpha\neq\beta),~~~\eta_{\alpha}^2=1,
\end{equation}
it leads to a representation of spin operators
\begin{equation}\label{eq:MF}
\sigma_{\alpha}=-i\sum_{\beta\gamma}\epsilon_{\alpha\beta\gamma}\eta_{\beta}\eta_{\gamma}.
\end{equation}

Noting the crucial property of the MFR
$\left\langle\sigma_{\alpha}(\tau)\sigma_{\beta}\right\rangle~=~\left\langle\eta_{\alpha}(\tau)\eta_{\beta}\right\rangle$ \cite{Mao.03.PRL,Shnirman.03.PRL,Schad.15.AP},
if we introduce the Keldysh GFs of Majorana-fermions
\begin{equation}\label{eq:mf_GF}
G_{\alpha\beta}(t,t^{\prime})~\equiv~-i\left\langle T_c~\eta_{\alpha}(t)\eta_{\beta}(t^{\prime})\right\rangle,
\end{equation}
then Keldysh GFs of spin operators can be rewritten in terms of the greater and lesser Keldysh GF of Majorana-fermions, namely,
$\Pi_{\alpha\beta}^<(t,t^{\prime}) = -G_{\alpha\beta}^<(t,t^{\prime}),
\Pi_{\alpha\beta}^>(t,t^{\prime}) = G_{\alpha\beta}^>(t,t^{\prime}),
\Pi_{\alpha\beta}^r(t,t^{\prime}) = \Theta(t-t^{\prime})[G_{\alpha\beta}^{>}(t,t^{\prime})+G_{\alpha\beta}^{<}(t,t^{\prime})],
\Pi_{\alpha\beta}^a(t,t^{\prime}) = -\Theta(t^{\prime}-t)[G_{\alpha\beta}^{>}(t,t^{\prime})+G_{\alpha\beta}^{<}(t,t^{\prime})]$.
Thus the evaluation of Keldysh GFs $\Pi_{\alpha\beta}$ turns into an evaluation of Keldysh GFs $G_{\alpha\beta}$. The latter enables a standard diagrammatic method. For later convenience, we denote $G_{\eta}\equiv G_{zz}$, $G_{\eta_x} \equiv G_{xx}$, $G_{\eta_y} \equiv G_{yy}$.

\subsection{Evaluation of Keldysh Green's functions}
By introducing the contour ordering, Keldysh GFs are formally and structurally equivalent to equilibrium counterparts \cite{Rammer.86.RMP,Haug.96.NULL}. So the Keldysh GF $G_{\eta}(t,t^{\prime})$ satisfies a Dyson-like equation
\begin{eqnarray}\label{eq:dyson}
G_{\eta}(t,t^{\prime}) &=& G_{\eta,0}(t,t^{\prime})+\int d t_1\int d t_2G_{\eta,0}(t,t_1)\cdot\nonumber\\
&&\Sigma_{\eta}(t_1,t_2)G_{\eta}(t_2,t^{\prime}),
\end{eqnarray}
where $G_{\eta,0}(t,t^{\prime})$ is the free Keldysh GF of Majorana-fermions, $\Sigma_{\eta}(t_1,t_2)$ corresponds to the self-energy due to the system-bath interaction. Once the information of $G_{\eta,0}(t,t^{\prime})$ and $\Sigma_{\eta}(t_1,t_2)$ are known, we can obtain $G_{\eta}(t,t^{\prime})$ via the above equation.

\subsubsection{Free Keldysh Green's functions}
We first consider free Keldysh GFs for Majorana-fermions. In the MFR, the transformed system Hamiltonian $H_s$ can be expressed as
\begin{equation}\label{eq:hs}
\tilde{H}_s~=~-i\frac{\varepsilon}{2}\eta_x\eta_y.
\end{equation}
Clearly we have $[\tilde{H}_s,\eta_z]=0$, so $\eta_z$ is time independent. We find in steady states $(\tau=t-t^{\prime})$ that
\begin{equation}
G_{\eta,0}^{r/a}(\tau)~=~\mp 2i \Theta(\pm \tau),
\end{equation}
which yields
\begin{equation}\label{eq:geta0_ra}
G_{\eta,0}^{r/a}(\omega)~=~\frac{2}{\omega\pm i\xi}
\end{equation}
with $\xi\rightarrow 0$ in the frequency domain.

We also need greater and lesser components of $G_{\eta_x,0}$ and $G_{\eta_y,0}$ in the calculations for self-energies below. From the Hamiltonian [Eq. (\ref{eq:hs})], equations of motion that $\eta_x$ and $\eta_y$ satisfy are given by $\eta_y(t)=\eta_y\cos\varepsilon t+\eta_x\sin\varepsilon t$ and
$\eta_x(t)=\eta_x\cos\varepsilon t-\eta_y\sin\varepsilon t$, respectively. Since $\varepsilon>0$, we choose the ground state, namely, the spin-down state as the initial condition such that
\begin{eqnarray}\label{eq:gxy}
G_{\eta_x,0}^>(\omega) &=& G_{\eta_y,0}^>(\omega) = -i2\pi\delta(\omega-\varepsilon),\nonumber\\
G_{\eta_x,0}^<(\omega) &=& G_{\eta_y,0}^<(\omega) = i2\pi\delta(\omega+\varepsilon),
\end{eqnarray}
It is worthwhile to mention that if we consider negative values of bias we should take the spin-up state (the corresponding ground state) as the initial state, but the resulting final expressions remain the same forms with those obtained below, implying that our results are symmetric functions of the bias.

We then turn to free Keldysh GFs for bath operators. According to Eq. (\ref{eq:t_interaction}), we introduce compact notations $B\equiv(\cos\Omega,\sin\Omega)^T$ and $B^{\dagger}\equiv(\cos\Omega,\sin\Omega)$ and define a matrix Keldysh GF for steady states
\begin{equation}
G_{B,0}(\tau)\equiv-i\left\langle T_c B(\tau)B^{\dagger}\right\rangle.
\end{equation}
Since the retarded and advanced components are totally determined by the lesser and greater ones, we only need
$G_{B,0}^>(\tau)= -i\left\langle B(\tau)B^{\dagger}\right\rangle$ and
$G_{B,0}^<(\tau)= -i\left\langle B^{\dagger}B(\tau)\right\rangle$. In order to simplify the calculation of matrix elements, we further introduce correlation functions $\Phi_{nm}(\tau)\equiv\frac{\Delta^2}{4}\left\langle e^{ni\Omega(\tau)}e^{mi\Omega}\right\rangle$ and $\tilde{\Phi}_{nm}(\tau)\equiv\frac{\Delta^2}{4}\left\langle e^{ni\Omega}e^{mi\Omega(\tau)}\right\rangle=\Phi_{nm}(-\tau)$ with $n,m=\pm$.
In terms of those correlation functions, we can rewrite matrix elements of $G_{B,0}^{>,<}$, for instance, we have
\begin{eqnarray}
\left\langle \cos\Omega(\tau)\cos\Omega\right\rangle &=& \frac{1}{\Delta^2}\left[\Phi_{++}(\tau)+\Phi_{+-}(\tau)\right.\nonumber\\
&&\left.+\Phi_{-+}(\tau)+\Phi_{--}(\tau)\right]
\end{eqnarray}
for an element of $G_{B,0}^>$. By replacing $\Phi_{nm}(\tau)$ with $\tilde{\Phi}_{nm}(\tau)$, we can obtain results for elements of $G_{B,0}^<$.

By assuming the two reservoirs are at their own thermal equilibrium states characterized by temperature $T_v$($v=L,R$), $\Phi_{nm}(\tau)$ and $\tilde{\Phi}_{nm}(\tau)$ can be evaluated by using the techniques of Feynman disentangling of operators \cite{Mahan.00.NULL}. Noting the addictive form of $\Omega$ for two baths [c.f., Eq. (\ref{eq:U})], we can directly write down final expressions according to the results of a single bath \cite{Liu.16.CP}
\begin{eqnarray}\label{eq:bath_corr}
\Phi_{+-}(\tau) &=& \frac{\Delta^2}{4}\exp\left[-\sum_{v=L,R}\left(Q_2^v(\tau)+i Q_1^v(\tau)\right)\right]\nonumber\\
&=& \Phi_{-+}(\tau)\\\label{eq:phipm}
\tilde{\Phi}_{+-}(\tau) &=& \frac{\Delta^2}{4}\exp\left[-\sum_{v=L,R}\left(Q_2^v(\tau)-i Q_1^v(\tau)\right)\right]\nonumber\\
&=& \tilde{\Phi}_{-+}(\tau)\label{eq:tphipm}
\end{eqnarray}
with
\begin{eqnarray}
Q_1^v(\tau) &=& \frac{2}{\pi}\int_0^{\infty}d\omega\frac{I_v(\omega)}{\omega^2}\sin\omega \tau,\label{eq:q1}\\
Q_2^v(\tau) &=& \frac{4}{\pi}\int_0^{\infty}d\omega\frac{I_v(\omega)}{\omega^2}\coth\left(\frac{\omega}{2T_v}\right)\sin^2\left(\frac{\omega \tau}{2}\right). \label{eq:q2}
\end{eqnarray}
Similarly, we can find $\Phi_{++}(\tau)=\Phi_{--}(\tau)$ and $\tilde{\Phi}_{++}(\tau)=\tilde{\Phi}_{--}(\tau)$.
By noting, for instance, $\left\langle \cos\Omega(\tau)\sin\Omega\right\rangle = \frac{i}{\Delta^2}\left[\Phi_{+-}(\tau)-\Phi_{++}(\tau)+\Phi_{--}(\tau)-\Phi_{-+}(\tau)\right]$, we deduce that the nondiagonal elements of $G_{B,0}^{>,<}$ are vanishing, so $G_{B,0}^{>,<}$ are actually diagonal matrices.

\subsubsection{Self-energy}
We now focus on the extraction of the fermionic self-energies. In the MFR, the interaction part can be rewritten as
\begin{equation}\label{eq:coupling}
\tilde{H}_I~\equiv~-i\frac{\Delta}{2}\left(\eta_y\eta_z\cos\Omega+\eta_z\eta_x\sin\Omega\right).
\end{equation}
For such a weak interaction, the spin dynamics should, in principle, be well described by the lowest-order self-energies. Noting the perturbation expansion for the Keldysh GFs is structurally equivalent to that of the equilibrium counterparts \cite{Rammer.86.RMP,Haug.96.NULL}, we can directly adopt the equilibrium result for the leading order self-energy $\Sigma_{\eta}$ \cite{Liu.16.CP} with equilibrium GFs being replaced by Keldysh GFs
\begin{eqnarray}\label{eq:se_t}
\Sigma_{\eta}(t_1,t_2) &=&  i\frac{\Delta^2}{4}\left[_{11}G_{B,0}(t_1,t_2)G_{\eta_y,0}(t_1,t_2)\right.\nonumber\\
&&\left.+_{22}G_{B,0}(t_1,t_2)G_{\eta_x,0}(t_1,t_2)\right]
\end{eqnarray}
with $_{11}G_{B,0}\equiv\left(\begin{array}{cc}
1 & 0
\end{array}
\right)G_{B,0}\left(\begin{array}{c}
1 \\
0
\end{array}
\right)$ and $_{22}G_{B,0}\equiv\left(\begin{array}{cc}
0 & 1
\end{array}
\right)G_{B,0}\left(\begin{array}{c}
0 \\
1
\end{array}
\right)$.

Using the Langreth theorem \cite{Haug.96.NULL} and expressions for the free Keldysh GFs, we find that
$\Sigma_{\eta}^>(\omega)=-i\Phi_{+-}(\omega-\varepsilon)$ and
$\Sigma_{\eta}^<(\omega)=i\tilde{\Phi}_{+-}(\omega+\varepsilon)$. Consequently, we have
\begin{equation}
\Sigma_{\eta}^r(\omega)-\Sigma_{\eta}^a(\omega) = -i\left(\Phi_{+-}(\omega-\varepsilon)+\tilde{\Phi}_{+-}(\omega+\varepsilon)\right),
\end{equation}
yielding
\begin{equation}\label{eq:eta_ra}
\mathrm{Im}\left[\Sigma_{\eta}^{r/a}(\omega)\right]~\equiv~\mp\frac{1}{2}\Gamma(\omega).
\end{equation}
with  $\Gamma(\omega)\equiv\left[\Phi_{+-}(\omega-\varepsilon)+\tilde{\Phi}_{+-}(\omega+\varepsilon)\right]$. In order to capture essential physics in the strong coupling regime, the real part $\Lambda$ of self-energies $\Sigma_{\eta}^{r/a}(\omega)$ should also be taken into account, as we have demonstrated for equilibrium systems. As determined by $\frac{1}{2}\left[\Sigma_{\eta}^r(\omega)+\Sigma_{\eta}^a(\omega)\right]$, its explicit form can be obtained in a similar way for equilibrium systems \cite{Liu.16.CP}
\begin{equation}\label{eq:eta_ra_real}
\Lambda(\omega) = \mathrm{Im}\left.\left[\tilde{\Phi}_{+-}(\lambda)\right]\right|_{\lambda=-i(\omega+\varepsilon)}+\mathrm{Im}\left.\left[\Phi_{+-}(\lambda)\right]\right|_{\lambda=-i(\omega-\varepsilon)},
\end{equation}
where $\Phi_{+-}(\lambda)$ and $\tilde{\Phi}_{+-}(\lambda)$ are bath correlations in the Laplace space. The sum of the retarded and advanced self-energies give the Keldysh component of self-energy
\begin{equation}\label{eq:sigma_k}
\Sigma_{\eta}^K(\omega) ~=~ i\left[\tilde{\Phi}_{+-}(\omega+\varepsilon)-\Phi_{+-}(\omega-\varepsilon)\right].
\end{equation}

\subsubsection{Expression for $\tilde{\chi}^{\prime\prime}_z(\omega)$}
Inserting free Keldysh GFs $G_{\eta,0}^{r/a}(\omega)$ and self-energies $\Sigma_{\eta}^{r/a}(\omega)$ into the Eq. (\ref{eq:dyson}), we have
\begin{equation}\label{eq:geta_ra}
G_{\eta}^{r/a}(\omega)
~=~\frac{2}{\omega-2\Lambda\pm i\Gamma}.
\end{equation}
The above results together with Eq. (\ref{eq:sigma_k}) lead to
\begin{equation}\label{eq:geta_k}
G_{\eta}^{K}(\omega)
~=~ \frac{4i \left[\tilde{\Phi}_{+-}(\omega+\varepsilon)-\Phi_{+-}(\omega-\varepsilon)\right]}{(\omega-2\Lambda)^2+ \Gamma^2}.
\end{equation}

Noting that $G^<=\frac{1}{2}(G^K-G^r+G^a)$, we obtain
\begin{equation}\label{eq:gzzl}
G_{\eta}^<(\omega)
~=~ \frac{4i \tilde{\Phi}_{+-}(\omega+\varepsilon)}{(\omega-2\Lambda)^2+ \Gamma^2}.
\end{equation}
Similarly,
\begin{equation}\label{eq:gzzg}
G_{\eta}^>(\omega)
~=~ -\frac{4i \Phi_{+-}(\omega-\varepsilon)}{(\omega-2\Lambda)^2+ \Gamma^2}.
\end{equation}
Since $\tilde{\chi}^{\prime\prime}_z(\omega)=\left[\Pi_{zz}^<(\omega)-\Pi_{zz}^>(\omega)\right]/2i$, we have
\begin{equation}\label{eq:chi}
\tilde{\chi}^{\prime\prime}_z(\omega)~=~2\frac{\Phi_{+-}(\omega-\varepsilon)-\tilde{\Phi}_{+-}(\omega+\varepsilon)}{(\omega-2\Lambda)^2+ \Gamma^2}.
\end{equation}
The above expression is one of the central results of this work.

\section{Ohmic dissipation}\label{sec:4}
In this section we shall apply the formal results of Eq. \ref{sec:3} to study the energy transfer for the case of Ohmic dissipation, that is, $s=1$ in the spectral density Eq. (\ref{eq:bath_spec}).

\subsection{Unbiased system}
We first consider unbiased spin systems with $\varepsilon=0$. In this situation, Eq. (\ref{eq:chi}) reduces to
\begin{equation}\label{eq:chi_O_unbias}
\tilde{\chi}^{\prime\prime}_z(\omega)~=~2\frac{\Phi_{+-}(\omega)-\tilde{\Phi}_{+-}(\omega)}{(\omega-2\Lambda)^2+\Gamma^2}
\end{equation}
with $2\Lambda$ and $\Gamma$ now the imaginary and real part of $ 2\left.\left[\Phi_{+-}(\lambda)+\tilde{\Phi}_{+-}(\lambda)\right]\right|_{\lambda=-i\omega}$, respectively. For later convenience, we introduce the correlation function of the $v$th bosonic bath
\begin{equation}\label{eq:cv}
C_v(\tau)~=~\frac{\Delta}{2}\exp\left[-Q_2^v(\tau)-i Q_1^v(\tau)\right],
\end{equation}
therefore the bath correlation functions can be expressed as
\begin{eqnarray}\label{eq:cLR}
\Phi_{+-}(\tau) &=& C_L(\tau)C_R(\tau),\nonumber\\
\tilde{\Phi}_{+-}(\tau) &=& C_L(-\tau)C_R(-\tau),
\end{eqnarray}
clearly, the two baths are involved non-additively.

In the weak coupling limit, we find (details can be found in Appendix \ref{sec:a1})
\begin{equation}\label{eq:chi_O_unbias_weak}
\tilde{\chi}^{\prime\prime}_z(\omega)\simeq 2\frac{\Delta^2\sum_vI_v(\omega)}{(\omega^2-\Delta^2)^2+\omega^2\left(\sum_vI_v(\omega)\coth\frac{\omega}{2T_v}\right)^2}.
\end{equation}
Inserting the above expression into Eq. (\ref{eq:heat_current}) yields
\begin{equation}\label{eq:hc_weak}
J_L~=~\frac{2}{\pi}\int_{0}^{\infty}d\omega\omega\frac{I_L(\omega)I_R(\omega)\Delta^2\left[n_L(\omega)-n_R(\omega)\right]}{\left(\omega^2-\Delta^2\right)^2+\omega^2\left(\sum_vI_v(\omega)\coth\frac{\omega}{2T_v}\right)^2},
\end{equation}
which is exactly the result obtained by a previous NEGF method \cite{Yang.14.EL}. When the bath temperature $T_{L/R}$ is comparable to or larger than the energy spacing $\Delta$ of the spin, the incoherent sequential process becomes the dominant heat transfer mechanism \cite{Ruokola.11.PRB}, thus the integrand of Eq. (\ref{eq:hc_weak}) with frequencies around $\Delta$ contributes the most to the heat current. As a result, Eq. (\ref{eq:hc_weak}) can be reduced to the result of QME\cite{Yang.14.EL}
\begin{equation}\label{eq:QME}
J_L~=~\Delta\frac{I_L(\Delta)I_R(\Delta)\left[n_L(\Delta)-n_R(\Delta)\right]}{I_L(\Delta)\left[2n_L(\Delta)+1\right]+I_R(\Delta)\left[2n_R(\Delta)+1\right]}.
\end{equation}
Thus our energy current formula can describe weak coupling regime, in contrast to the NE-NIBA whose result \cite{Nicolin.11.JCP,Nicolin.11.PRB}
\begin{equation}\label{eq:NIBA}
J_L~=~\frac{1}{4\pi}\int_{-\infty}^{\infty}\omega\left[C_R(\omega)C_L(-\omega)-C_R(-\omega)C_L(\omega)\right]d\omega
\end{equation}
can only be applied to the strong coupling regime \cite{Chen.13.PRB}.

\begin{figure}[tbh]
  \centering
  \includegraphics[width=0.95\columnwidth]{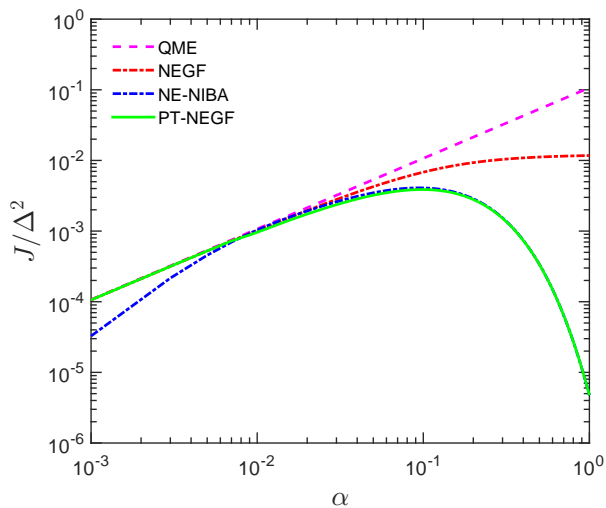}
\caption{(Color Online)Comparison between theoretical results on the energy current. The pink dashed line denotes result of the QME [Eq. (\ref{eq:QME})], the blue dashed-dotted line denotes the result of the NE-NIBA [Eq. (\ref{eq:NIBA})], the red dashed-dotted line denotes the result of a previous NEGF method [Eq. (\ref{eq:hc_weak})] and the green solid line denotes the result of the present PT-NEGF method [Eq. (\ref{eq:heat_current}) together with Eq.(\ref{eq:chi_O_unbias})]. Parameters are $T_L/\Delta=1.4$, $T_R/\Delta=1.2$, $\alpha_L=\alpha_R=\alpha/2$, $\omega_c/\Delta=30$ such that the nonadiabatic limit is fulfilled.}
\label{fig:hc_unbias}
\end{figure}

In order to see the performance of our result in a wide range of the coupling strength, a comparison between various theoretical predictions for the energy current is shown in Fig. \ref{fig:hc_unbias}.
From the figure, it is expected that our formula matches the results of the QME and the NEGF in the weak coupling regime, while the NE-NIBA underestimates values of the heat current. As the coupling strength increases, our result depicts a "turn-over" behavior which is in accord with exact numerical results \cite{Velizhanin.08.CPL} as well as the NE-NIBA's prediction, thus our result is distinct from results of the QME and NEGF. This "turn-over" phenomenon results from a renormalization effect of tunneling between two spin states in the strong coupling regime. In the strong coupling regime, the profile of our formula almost coincides with the result of the NE-NIBA. We attribute this agreement to the fact that approximations underlying our theory bear a close resemblance to that of the NIBA framework for unbiased spin systems as has been noted in equilibrium cases \cite{Liu.16.CP}. These features indicate that our theory indeed provides a comprehensive and unified interpretation for energy transfer over a wide range of the coupling strength, a considerable improvement over existing theories.

We next consider the influence of the energy spacing $\Delta$ of the
spin system on the energy conductance. According to the definition, $\kappa$ is totally determined by $\left.\tilde{\chi}^{\prime\prime}_z(\omega)\right|_{T_L= T_R=T}$, in order to evaluate it, we only need Laplace transforms of bath correlations [Eq. (\ref{eq:bath_corr})] at equilibrium states (see details in Appendix \ref{sec:a2}). In the scaling limit of $\omega_c/T\gg1$, we already know that \cite{Liu.16.CP}
\begin{eqnarray}\label{eq:bc_lambda}
\Phi_{+-}(\lambda) &=& \frac{\Delta^2}{4\omega_c}e^{-i\pi\alpha}\frac{\Gamma(1-2\alpha)\Gamma(\alpha+\lambda/2\pi T)}{\Gamma(1-\alpha+\lambda/2\pi T)}\left(\frac{2\pi T}{\omega_c}\right)^{2\alpha-1}\nonumber\\
&=& \tilde{\Phi}_{+-}^{\ast}(\lambda).
\end{eqnarray}
\begin{figure}[tbh]
  \centering
  \includegraphics[width=1.\columnwidth]{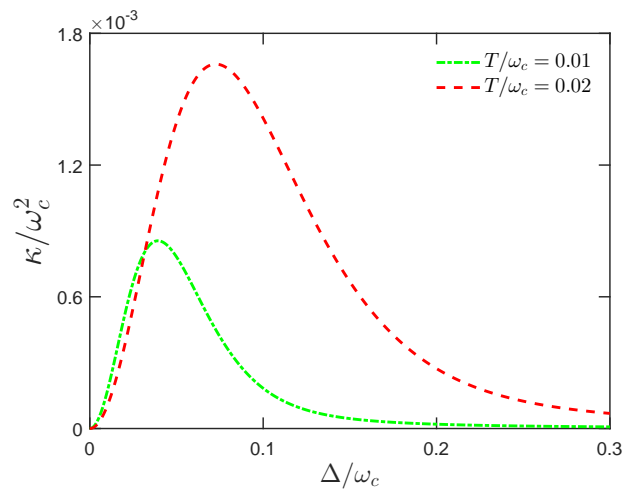}
\caption{(Color Online)Dependence of the energy conductance on the energy spacing $\Delta$ of the spin system for different temperatures. We choose $\alpha=0.1$, $T_L=T_R=T$.}
\label{fig:kappa_spacing}
\end{figure}
Results for $\kappa$ depicted in Fig. \ref{fig:kappa_spacing} show a "turn-over" behavior as a
function of the energy spacing $\Delta$, which is in accordance to the findings of the multilayer
multiconfiguration time-dependent Hartree theory \cite{Velizhanin.08.CPL} as well as the influence functional path integral method \cite{Boudjada.14.JPCA}. This dependence is due to the resonant character of energy transfer. The energy transfer is most efficient if the energy scale of the bridge subsystem is comparable to the temperature of the bath, as can be found that the value of $\Delta$ corresponding to the "turn-over" point for $T/\omega_c=0.02$ is almost double that for $T/\omega_c=0.01$.

\subsection{Biased system}
In this subsection we turn to the biased spin system. Although its dissipative dynamics has been extensively studied \cite{Leggett.87.RMP,Weiss.12.NULL,Weiss.89.PRL,Goerlich.89.EL,Weiss.86.EL,Grabert.85.PRL,Fisher.85.PRL}, a thorough understanding of its energy transfer characteristic limits to the time-dependent case \cite{Carrega.16.PRL}. So far, no numerical methods could address effects of a finite bias at steady states. On the theoretical side, only the QME and NE-NIBA can involve bias in their formula \cite{Boudjada.14.JPCA}. However, their validity regime are limited. It is then of necessity and interest to explore the energy transfer behaviors of the NESB model with a finite bias using our approach.

For biased systems, the QME framework predicted that \cite{Boudjada.14.JPCA}
\begin{equation}\label{eq:kq}
J_{Q}~=~\frac{\Delta^2}{\omega_0}\frac{I_L(\omega_0)I_R(\omega_0)[n_L(\omega_0)-n_R(\omega_0)]}{I_L(\omega_0)[1+2n_L(\omega_0)]+I_R(\omega_0)[1+2n_R(\omega_0)]}
\end{equation}
with $\omega_0\equiv\sqrt{\Delta^2+\varepsilon^2}$ the energy spacing of the spin. While the energy current of the NE-NIBA reads \cite{Nicolin.11.JCP,Nicolin.11.PRB}
\begin{eqnarray}\label{eq:kn}
J_N &=& \frac{1}{2\pi}\int_{-\infty}^{+\infty}\omega d\omega\left[P_1C_R(\omega)C_L(\varepsilon-\omega)\right.\nonumber\\
&&-\left.P_0C_R(-\omega)C_L(\omega-\varepsilon)\right],
\end{eqnarray}
where $C_v(\omega)$ is the Fourier transform of equation (\ref{eq:cv}), $P_{0,1}$ denote the steady-state population of the spin states which are determined by the Fourier transform of bath correlation functions [Eq. (\ref{eq:cLR})]:
\begin{eqnarray}
P_0 &=& \frac{\Phi_{+-}(\varepsilon)}{\Phi_{+-}(\varepsilon)+\tilde{\Phi}_{+-}(\varepsilon)},\nonumber\\
P_1 &=& \frac{\tilde{\Phi}_{+-}(\varepsilon)}{\Phi_{+-}(\varepsilon)+\tilde{\Phi}_{+-}(\varepsilon)}.
\end{eqnarray}

\begin{figure}[tbh]
  \centering
  \includegraphics[width=1.\columnwidth]{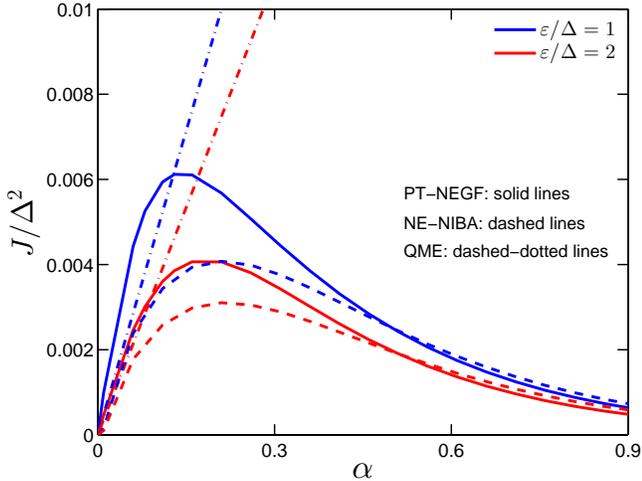}
\caption{(Color Online)Behaviors of the energy current with varying bias (blue denotes $\varepsilon=1$, red denotes $\varepsilon=2$) as a function of coupling strength. We compare PT-NEGF results (solid lines) to QME [Eq. (\ref{eq:kq})] (dashed-dotted line) and NE-NIBA [Eq. (\ref{eq:kn})] (dashed lines). The inset shows the dependence of heat conductance on the bias for different coupling strengths predicted by the PT-NEGF. We choose $T_L/\Delta=1.4$, $T_R/\Delta=1.2$, $\alpha_L=\alpha_R=\alpha/2$, $\omega_c/\Delta=30$ such that the nonadiabatic limit is fulfilled.}
\label{fig:hc_bias}
\end{figure}

Results for the energy current are shown in Fig. \ref{fig:hc_bias}. From the figure,
we found that the PT-NEGF scheme still agree with the QME in the weak coupling regime. Increasing the coupling strength, the QME predicts an almost linearly increasing $\kappa$ which is qualitatively incorrect, whereas our results and the NE-NIBA show turn-over behaviors, although the turn-over point differs. In the strong coupling regime, the PT-NEGF approaches the NE-NIBA. Thus the PT-NEGF can describe the energy transfer in biased systems.

\begin{figure}[tbh]
  \centering
  \includegraphics[width=1.\columnwidth]{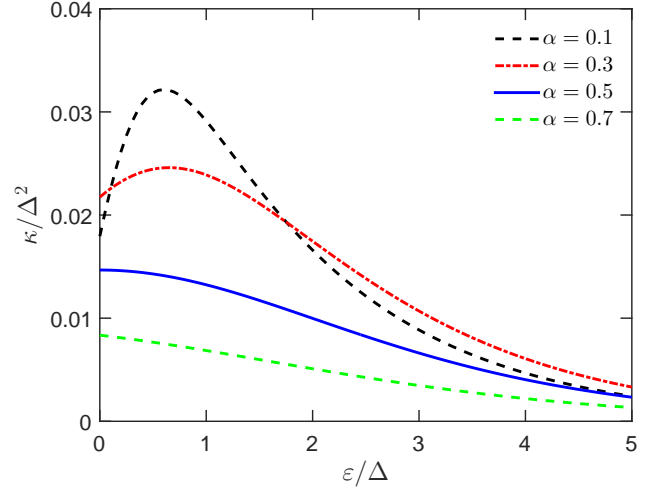}
\caption{(Color Online)The dependence of energy conductance on the bias for different coupling strengths predicted by the PT-NEGF. We choose $T_L/\Delta=T_R/\Delta=1.2$, $\omega_c/\Delta=30$ such that the nonadiabatic limit is fulfilled.}
\label{fig:kappa_bias_alpha}
\end{figure}

In the intermediate coupling regime, we further notice an interesting phenomenon that the energy conductance is a nonmonotonic function of the bias as can be seen from Fig. \ref{fig:kappa_bias_alpha}.
Such a dependence is directly in contrast to the NE-NIBA's prediction $\kappa\propto\varepsilon/\sinh(\varepsilon/T)$ for $\alpha=0.1-0.5$ \cite{Boudjada.14.JPCA}. Noted that for the spin system with fixed $\Delta$, the energy spacing $\omega_0$ increases as bias increases, so the nonmonotonic dependence of $\kappa$ on $\varepsilon$ at moderate couplings is similar to the results presented in Fig. \ref{fig:kappa_spacing}. Therefore we can attribute this nonmonotonic behavior to the energy resonance between the bath and the spin. The NE-NIBA fails to capture this behavior, thus it is invalid in the intermediate coupling regime in the presence of bias, only in the strong coupling regime can we observe predicted monotonic behaviors of $\kappa$ as can be seen from the figure.

\begin{figure}[tbh]
  \centering
  \includegraphics[width=1.\columnwidth]{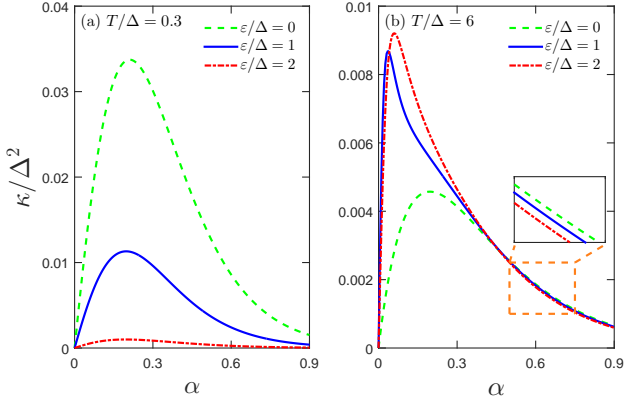}
\caption{(Color Online)Behaviors of the heat conductance with varying bias as a function of coupling strength for (a)$T/\Delta=0.3$, (b)$T/\Delta=6$, the inset shows details of $\kappa$ in the strong coupling regime. We choose $T_L=T_R=T$ and $\omega_c/T=30$ such that the nonadiabatic limit is fulfilled.}
\label{fig:kappa_bias}
\end{figure}

In accordance with such a nonmonotonic dependence, we would expect that if $\omega_0$ is always smaller (larger) than the bath temperature $T$, $\kappa$ should be a monotonically increasing (decreasing) function of $\varepsilon$ at moderate couplings. To verify this, we choose appropriate temperatures and vary bias. Results are shown in Fig. \ref{fig:kappa_bias}. As can be seen from the figure, the dependence of $\kappa$ on $\varepsilon$ in the intermediate coupling regime indeed meets our expectation. We also observe that $\kappa$ is always a monotonically decreasing function of $\varepsilon$ in the strong coupling regime, regardless of values of temperature, since this regime is dominated by the strong dissipation from the bath. Noting values of the bias can be adjusted by changing the applied magnetic field, the interplay of the temperature, the bias and the coupling strength may offer a nontrivial quantum control knob over heat transfer at the nanoscale, our future work will address this aspect in more details.

\section{Summary}\label{sec:5}
Methodologically, we established a polaron-transformed nonequilibrium Green's function method (PT-NEGF) to investigate energy transfer in nonequilibrium spin-boson models. In contrast to previous NEGF methods treating the system-bath interaction as a perturbation, our NEGF method utilizes a polaron transformation and takes the tunneling term as a perturbation. Furthermore, in order to evaluate terms in the expansion series, we adopt the Majorana-fermion representation such that standard Feynman diagram techniques as well as the Dyson's equation can be applied to spin systems. By doing so, our theoretical scheme goes beyond existing methods, it can tackle the strong coupling regime and include effects of a finite bias.

To demonstrate the utility of the approach, we first consider unbiased spin systems. Our analysis shows that the PT-NEGF method can give a comprehensive and unified interpretation for energy transfer over wide ranges of the coupling strength, a considerable improvement over existing theories. The predicted behavior of energy conductance as a function of the energy spacing of the spin is in accord with exact numerical simulations. When a finite bias plays a role, we found the energy conductance endows a nonmonotonic bias dependence at moderate coupling strengths which is not reported in present literatures. We attribute this phenomenon to the resonant character of energy transfer in such systems. To verify our interpretation, we further consider different temperature regimes and obtain self-consistent results. These features thus make our theory stand out from previous considerations.

By noting the interplay of the coupling strength, the bias and the temperature, we will consider how to use quantum control to optimize the energy transfer process. Our future work will also address energy transfer behaviors in the vicinity of a dissipative quantum phase transition point at very low temperatures \cite{Chung.09.PRL}.

\begin{acknowledgments}
The authors thank J. Cao, H. Zhou and J. Ren for highly useful discussions. Support from the National Nature Science Foundation of China with Grant No. 11574050 and the National Basic Research Program of China with Grant No. 2012CB921401 are gratefully acknowledged.
\end{acknowledgments}

\appendix
\section{Weak coupling limit of equation (\ref{eq:chi_O_unbias})}\label{sec:a1}
In the nonadiabatic limit of $\omega_c/T_v\gg 1$, Eq. (\ref{eq:cv}) has a following explicit form \cite{Dattagupta.89.JPCM,Weiss.12.NULL,Gorlich.88.PRB}
\begin{equation}
C_v(\tau)~=~\frac{\Delta}{2}\mathrm{exp}\left[-i\pi\alpha_v\, \mathrm{sgn}(\tau)\right]\left(\frac{\pi T_v}{\omega_c\sinh(\pi T_v|\tau|)}\right)^{2\alpha_v}.
\end{equation}
Taking the weak coupling limit, we only need to keep the leading order of $C_v(\tau)$ as $O(\alpha_v)$ such that Eq. (\ref{eq:cLR}) becomes
\begin{eqnarray}
\Phi_{+-}(\tau) &\approx& \frac{\Delta^2}{4}\left(1-\sum_v\alpha_v\eta_v\right),\nonumber\\
\tilde{\Phi}_{+-}(\tau) &\approx& \frac{\Delta^2}{4}\left(1-\sum_v\alpha_v\eta_v^{\ast}\right).
\end{eqnarray}
with
\begin{equation}
\eta_v=i\pi\mathrm{sgn}(\tau)+2\ln\frac{\omega_c\sinh(\pi T_v|\tau|)}{\pi T_v}
\end{equation}
and $\eta_v^{\ast}$ its complex conjugate. By applying the Laplace transform, we have
\begin{widetext}
\begin{equation}
\left.\left[\Phi_{+-}(\lambda)+\tilde{\Phi}_{+-}(\lambda)\right]\right|_{\lambda=-i\omega} ~=~ \frac{i\Delta^2}{2\omega}+\frac{i\Delta^2}{\omega}\sum_v\alpha_v\left[-i\pi\coth\frac{\omega}{2T_v}+\psi\left(\frac{i\omega}{2\pi T_v}\right)+\frac{\pi T_v}{i\omega}-\ln\frac{\omega_c}{2\pi T_v}+\gamma_E\right]
\end{equation}
\end{widetext}
with $\gamma_E$ the Euler-Mascheroni constant $0.577216\cdots$ and $\psi(z)$ the digamma function. Therefore, $2\Lambda$ and $\Gamma$ can be expressed as
\begin{eqnarray}\label{eq:LG_J}
2\Lambda &\simeq& \frac{\Delta^2}{\omega},\nonumber\\
\Gamma &=& \frac{\Delta^2}{\omega^2}\sum_vI_v\coth\frac{\omega}{2T_v}.
\end{eqnarray}
In deriving $\Gamma$, we have used the fact that
\begin{eqnarray}\label{eq:im_digamma}
2i\mathrm{Im}\psi\left(\frac{i\omega}{2\pi T_v}\right) &=& \psi\left(\frac{i\omega}{2\pi T_v}\right)-\psi\left(-\frac{i\omega}{2\pi T_v}\right)\nonumber\\
&=& i\pi\coth\frac{\omega}{2T_v}-\frac{2\pi T_v}{i\omega}.
\end{eqnarray}
Similarly, we have
\begin{eqnarray}\label{eq:pp_J}
\Phi_{+-}(\omega)-\tilde{\Phi}_{+-}(\omega) &=& 2 \mathrm{Re} \left.\left[\Phi_{+-}(\lambda)-\tilde{\Phi}_{+-}(\lambda)\right]\right|_{\lambda=i\omega}\nonumber\\
&=& \frac{\Delta^2}{\omega^2}\sum_vI_v,
\end{eqnarray}
where $"\mathrm{Re}"$ denotes the real part. Inserting Eqs. (\ref{eq:LG_J}) and (\ref{eq:pp_J}) into the expression of $\tilde{\chi}^{\prime\prime}_z(\omega)$ [Eq. (\ref{eq:chi_O_unbias})] yields Eq. (\ref{eq:chi_O_unbias_weak}) in the main text.

\section{Determining $\tilde{\chi}^{\prime\prime}_z(\omega)$}\label{sec:a2}
In this section we show that $\tilde{\chi}^{\prime\prime}_z(\omega)$ can be fully evaluated with the knowledge of the Laplace transform of bath correlation functions, namely, $\Phi_{+-}(\lambda)$ and $\tilde{\Phi}_{+-}(\lambda)$. First of all, we have
\begin{eqnarray}\label{eq:a2_g}
\Gamma(\omega) &=& \left[\Phi_{+-}(\omega-\varepsilon)+\tilde{\Phi}_{+-}(\omega+\varepsilon)\right]\nonumber\\ &=& 2\mathrm{Re}\left.\left[\tilde{\Phi}_{+-}(\lambda)\right]\right|_{\lambda=-i(\omega+\varepsilon)}\nonumber\\
&&+2\mathrm{Re}\left.\left[\Phi_{+-}(\lambda)\right]\right|_{\lambda=-i(\omega-\varepsilon)}
\end{eqnarray}
. Similarly, we find
\begin{eqnarray}\label{eq:a2_p}
\Phi_{+-}(\omega-\varepsilon)-\tilde{\Phi}_{+-}(\omega+\varepsilon) &=& 2\mathrm{Re}\left.\left[\Phi_{+-}(\lambda)\right]\right|_{\lambda=-i(\omega-\varepsilon)}\nonumber\\
&&-2\mathrm{Re}\left.\left[\tilde{\Phi}_{+-}(\lambda)\right]\right|_{\lambda=-i(\omega+\varepsilon)}.
\end{eqnarray}
The above equations together with Eq. (\ref{eq:eta_ra_real}) lead to $\tilde{\chi}^{\prime\prime}_z(\omega)$ [Eq. (\ref{eq:chi})].

%\bibliography{a}

\begin{thebibliography}{60}
\bibitem{Koh.07.PRB}Y. K. Koh and D. G. Cahill, Phys. Rev. B {\bf76}, 075207(2007).
\bibitem{Lefevre.05.PSI}S. Lef\'evre and S. Volz, Phys. Sci. Instrum. {\bf76}, 033701(2005).
\bibitem{Chavez.14.APLM}E. Ch\'avez-\'Angel et. al., Appl. Phys. Lett. Mater. {\bf2}, 012113(2014).
\bibitem{Reparaz.14.PSI}J. Reparaz et. al., Phys. Sci. Instrum. {\bf85}, 034901(2014).
\bibitem{Wang.08.EPJB}J.-S. Wang, J. Wang, and J. T. Lu, Eur. Phys. J. B {\bf62}, 381(2008).
\bibitem{Dubi.11.RMP}Y. Dubi and M. Di Ventra, Rev. Mod. Phys. {\bf83}, 131(2011).
\bibitem{Segal.05.PRL}D. Segal and A. Nitzan, Phys. Rev. Lett. {\bf94}, 034301(2005).
\bibitem{Ren.10.PRL}J. Ren, P. H\"anggi, and B. Li, Phys. Rev. Lett. {\bf104}, 170601(2010).
\bibitem{Boudjada.14.JPCA}N. Boudjada and D. Segal, J. Phys. Chem. A {\bf118}, 11323(2014).
\bibitem{Segal.06.PRB}D. Segal, Phys. Rev. B {\bf73}, 205415(2006).
\bibitem{Ruokola.11.PRB}T. Ruokola and T. Ojanen, Phys. Rev. B {\bf83}, 045417(2011).
\bibitem{Thingna.14.JCP}J. Thingna, H. Zhou, and J.-S. Wang, J. Chem. Phys. {\bf141}, 194101(2014).
\bibitem{Velizhanin.10.JCP}K. A. Velizhanin, M. Thoss, and H. Wang, J. Chem. Phys. {\bf133}, 084503 (2010).
\bibitem{Wang.13.FP}J.-S.Wang, B. K. Agarwalla, H. Li, and J. Thingna, Front. Phys. {\bf9}, 673(2013).
\bibitem{Yang.14.EL}Y. Yang and C. Wu, Europhys. Lett. {\bf107}, 30003(2014).
\bibitem{Mao.03.PRL}W. Mao, P. Coleman, C. Hooley, and D. Langreth, Phys. Rev. Lett. {\bf91}, 207203(2003).
\bibitem{Shnirman.03.PRL}A. Shnirman and Y. Makhlin, Phys. Rev. Lett. {\bf91}, 207204(2003).
\bibitem{Florens.11.PRB}S. Florens, A. Freyn, D. Venturelli, and R. Narayanan, Phys. Rev. B {\bf84}, 155110(2011).
\bibitem{Schad.15.AP}P. Schad, Y.Makhlin, B. Narozhny, G. Sch\"on, and A. Shnirman, Ann. Phys. {\bf361}, 401 (2015).
\bibitem{Nicolin.11.JCP}L. Nicolin and D. Segal, J. Chem. Phys. {\bf135}, 164106(2011).
\bibitem{Nicolin.11.PRB}L. Nicolin and D. Segal, Phys. Rev. B {\bf84}, 161414(2011).
\bibitem{Wang.15.SR}C. Wang, J. Ren, and J. Cao, Sci. Rep. {\bf5}, 11787(2015).
\bibitem{Chen.13.PRB}T. Chen, X.-B. Wang, and J. Ren, Phys. Rev. B {\bf87}, 144303(2013).
\bibitem{Velizhanin.08.CPL}K. A. Velizhanin, H. Wang, and M. Thoss, Chem. Phys. Lett. {\bf460}, 325(2008).
\bibitem{Segal.13.PRB}D. Segal, Phys. Rev. B {\bf87}, 195436 (2013).
\bibitem{Saito.13.PRL}K. Saito and T. Kato, Phys. Rev. Lett. {\bf111}, 214301(2013).
\bibitem{Carrega.16.PRL}M. Carrega, P. Solinas, M. Sassetti, and U. Weiss, Phys. Rev. Lett. {\bf116}, 240403(2016).
\bibitem{Makarov.94.CPL}D. E. Makarov and N. Makri, Chem. Phys. Lett. {\bf221}, 482(1994).
\bibitem{Bijay.16.NULL}B. K. Agarwalla and D. Segal, arXiv:1612.01008.
\bibitem{Weiss.12.NULL}U. Weiss, {\it Quantum Dissipative Systems}(World Scientific, Singapore, 2012).
\bibitem{Liu.16.CP}J. Liu, H. Xu, B. Li, and C. Wu, Chem. Phys. (2016), in press.
\bibitem{Leggett.87.RMP}A. J. Leggett, S. Chakravarty, A. T. Dorsey, M. P. A. Fisher, A. Garg, and W. Zwerger, Rev. Mod. Phys. {\bf59}, 1(1987).
\bibitem{Esposito.15.PRL}M. Esposito, M. A. Ochoa, and M. Galperin, Phys. Rev. Lett. {\bf114}, 080602(2015).
\bibitem{Rammer.86.RMP}J. Rammer and H. Smith, Rev. Mod. Phys. {\bf58}, 323(1986).
\bibitem{Haug.96.NULL}H. Haug and A. P. Jauho, {\it Quantum Kinetics in Transport and Optics of Semiconductors}(Springer-Verlag, Berlin, 1996).
\bibitem{Ojanen.08.PRL}T. Ojanen and A.-P. Jauho, Phys. Rev. Lett. {\bf100}, 155902(2008).
\bibitem{Schad.16.PRB}P. Schad, A. Shnirman, and Y. Makhlin, Phys. Rev. B {\bf93}, 174420(2016).
\bibitem{Lee.12.JCP}C. K. Lee, J. Moix, and J. Cao, J. Chem. Phys. {\bf136}, 204120(2012).
\bibitem{Tsvelik.92.PRL}A. M. Tsvelik, Phys. Rev. Lett. {\bf69}, 2142(1992).
\bibitem{Mahan.00.NULL}G. D. Mahan, {\it Many-Particle Physics}(Plenum, New York, 2000).
\bibitem{Weiss.89.PRL}U. Weiss and M. Wollensak, Phys. Rev. Lett. {\bf62}, 1663(1989).
\bibitem{Goerlich.89.EL}R. G\"orlich, M. Sassetti, and U. Weiss, Europhys. Lett. {\bf10}, 507(1989).
\bibitem{Weiss.86.EL}U. Weiss and H. Grabert, Europhys. Lett. {\bf2}, 667(1986).
\bibitem{Grabert.85.PRL}H. Grabert and U. Weiss, Phys. Rev. Lett. {\bf54}, 1605(1985).
\bibitem{Fisher.85.PRL}M. P. A. Fisher and A. T. Dorsey, Phys. Rev. Lett. {\bf54}, 1609(1985)
\bibitem{Chung.09.PRL}C.-H. Chung, K. Le Hur, M. Vojta, and P. W\"olfle, Phys. Rev. Lett. {\bf102}, 216803(2009).
\bibitem{Dattagupta.89.JPCM}S. Dattagupta, H. Grabert, and R. Jung, J. Phys.: Condens. Matter {\bf1}, 1405 (1989).
\bibitem{Gorlich.88.PRB}R. G\"orlich and U. Weiss, Phys. Rev. B {\bf38}, 5245(1988).
\end{thebibliography}

\end{document}